\documentclass[aps,prl,twocolumn,showpacs,superscriptaddress,groupedaddress]{revtex4-1}

\DeclareUnicodeCharacter{0301}{\'{e}}
\usepackage{graphicx}
\usepackage{hyperref}
\usepackage{xcolor}
\usepackage{color}
\usepackage{amsmath}
\usepackage{mathrsfs}
\usepackage{amsfonts}
\usepackage{amssymb}
\usepackage{varioref}
\usepackage{float}
\usepackage{dcolumn} 
\usepackage{bm}
\usepackage{relsize}
\usepackage[flushleft]{threeparttable}
\newcommand{\cblue}{\color{blue}}

\graphicspath{{./plots/}} 
\begin{document}
\title{ Compression of a confined semiflexible polymer under direct and oscillating fields}
\author{Keerthi Radhakrishnan}
\email{keerthirk@iiserb.ac.in}
\author{Sunil P Singh}
\email{spsingh@iiserb.ac.in}
\affiliation{Department of Physics, Indian Institute Of Science Education and Research, Bhopal 462 066, Madhya Pradesh, India}
\date{\today}

\begin{abstract}

 { The folding transition of biopolymers from the coil to compact structures has attracted wide research interest in the past and is well studied  in polymer physics. Recent seminal works on DNA in confined devices have shown that these long biopolymers tend to collapse under an external field, contrary to the previously reported stretching.  These long folded structures have a tendency to form
knots that has profound implications in gene regulation and various other biological functions.  These knots have been mechanically induced via optical tweezers, nano-channel confinement, etc., until recently, where uniform field driven compression lead to self entanglement of DNA. In this work, we capture the compression of a confined semiflexible polymer under direct and oscillating fields, using a coarse-grained computer simulation model in the presence of long-range hydrodynamics. Within this framework, we show that subjected to direct field, chains in stronger confinements exhibit substantial compaction, contrary to the one in moderate confinements or bulk, where such compaction is absent. Interestingly, an  alternating field within an optimum frequency can effectuate this compression even in moderate or no confinement. Additionally, we show that the bending rigidity has a profound influence on the chain’s folding favourability under direct and alternating fields.  This field induced collapse is a quintessential hydrodynamic phenomenon, resulting in intertwined knotted structures, even for shorter chains, unlike DNA knotting experiments, where it happens exclusively for longer chains.
 }

\end{abstract}
\maketitle

%

{
}

\section{INTRODUCTION}
Nano-fluidic devices have been widely used for the characterization of macromolecules like  separation\cite{Han_Sc_2000,Muthu_Book_2011},   and sequencing of DNA\cite{Han_Sc_2000,Muthu_Book_2011,branton2008potential,benner2007sequence,chen2004probing,vodyanoy1992sizing,kasianowicz1996characterization}, protein synthesis, transport of biopolymers, etc.\cite{kong2002modeling,venkatesan2011nanopore,derrington2010nanopore,branton2008potential,benner2007sequence,chen2004probing,vodyanoy1992sizing,kasianowicz1996characterization,jeon2014polymer,wong2010polymer} The fluid flow in such narrow capillaries profoundly influences the structural \cite{Han_Sc_2000,Muthu_Book_2011,jonsson1993orientation,netz_prl_2003,Singh_2014_JCP,odijk2008_PRE_scaling,muralidhar2014backfolding} and   dynamical  behavior of soft-deformable macromolecules\cite{Stein_PNAS_2006,Jendrejack_JCP_2004,jendrejack2003dna, Chelakkot_EPL_2010,Steinhauser_ACS_2012,hickey2013electrophoretic,Han_Sc_2000,Muthu_Book_2011,branton2008potential,benner2007sequence,chen2004probing,vodyanoy1992sizing,kasianowicz1996characterization}.  Few recent experiments have reported compression of confined long DNA strand into isotropic globules under uniform electric field\cite{Jing_PNAS_2011,zhou_2015_collapse,Zhou_PRL_2011}, despite  previously reported stretching of DNA seen under field\cite{jonsson1993orientation,ueda1998electrophoresis,ueda1999dynamics,kaji2003stretching,liu2010molecular,hsiao2011unfolding,netz_prl_2003,Frank_JCP_2009,netz2003polyelectrolytes}.  Similar shrinkage of a confined polyelectrolyte chain has been also captured in simulations\cite{radhakrishnan2021collapse} under AC field, with the exception that the shrinkage was manifested as backfoldings rather than an isotropic compression. These works substantiate the inevitability of hydrodynamic interactions in inducing such compression. Similar, hydrodynamic flow induced structural shrinkage is seen across   diverse soft-matter systems like U-shape bending of elastic rods under field\cite{lagomarsino2005hydro_Ushape}, shear induced compaction\cite{Singh_2018_JCP,sendner2009single}, compaction  of short chains under sedimenting fields\cite{Netz_PRL_2009, schlagberger2008anomalous}, etc. 



One direct repercussion of the chain compaction  is the enhanced favourability of self knotting observed within these structures. The formation of these spontaneous knots along a long polymer chain like DNA\cite{rybenkov1993probability,bao2003behavior,deibler2001topoisomerase,arai1999tying} and other proteins\cite{taylor2003protein,virnau2006intricate} is a recurrent phenomenon in biological processes.
 The use of chain compaction either by spatial confinement, compaction against slit barriers\cite{amin2018nanofluidic}, or molecular crowding\cite{d2015molecular} has been deemed extremely useful in effectuating these knottings.  The use of direct or  oscillatory fields in dielectrophoresis is another important protocol in obtaining a spectra of such self entangled structures. 

In this article, we elucidate how a confined generic semiflexible chain exhibits large-scale  compression under both direct and oscillating  uniform fields. The course of investigation is mostly steered along deciphering the sensitiveness of this  intriguing phenomenon to varying physical complexities like bending rigidity\cite{matthews2012effect,orlandini2005entanglement,zhu2021revisiting}, degree of confinement\cite{nakajima2013localization,micheletti2012knotting,reifenberger2015topological}, and chain length, along with plausible driving mechanism which remains elusive so far. While hydrodynamic flow field remains precursor to such a phenomenon, the above factors also acts as important ancillaries in dictating the folding favourability.

\section{MODEL}
For this we model a semiflexible polymer along with solvent molecules confined in a cylindrical tube, periodic along  its axis. 
The polymer is modelled as a bead-spring chain, which consists of $N_m$ monomers. The bond connectivity between adjacent monomers is ensured using a harmonic spring potential  $
 U_{s} = \sum_{i=1}^{N_m-1}\frac{k_s}{2} \left( { r}_{i,i+1} -l_0 \right)^2$, where $l_0$, $k_s$,  ${r}_{i,i+1} =
|{\bm r}_{i+1} - {\bm r}_i|$, and ${\bf r}_i$ denotes the equilibrium bond length, the spring constant,  magnitude of the bond vector, and  position of the $i$th monomer, respectively. The semi-flexibility is introduced via a bending potential, 
$ U_b = \sum_{i=1}^{N_m-2}\frac{k_b}{2} \left( { r}_{i,i+1} - r_{i+1,i+2}\right)^2$, where $k_b$ is the bending rigidity that dictates  the stiffness or persistence length of the chain.

 Furthermore, excluded volume interactions among chain  monomers  are incorporated via  standard repulsive shifted  Lennard-Jones (LJ) potential. 
 The solvent molecules are  modelled using multi-particle collision dynamics\cite{Kapral_ACP_2008,Gompper_APS_2009}, which is a particle based mesoscopic simulation technique that incorporates both hydrodynamic interactions and thermal fluctuations.  
 The elaborate description of the method can be found in references cited  herein \cite{Kapral_ACP_2008,Gompper_APS_2009,Ripoll04,Malevanets_MSM_1999,Lamura_EPL_2001,Singh_2014_JCP}.
 
 The interaction of monomers with wall is considered via a similar LJ potential described earlier, where  a monomer feels a repulsive force within a distance $R_c \leq 2^{1/6}\sigma/2$ from the wall. 
 An external field $G$ is imposed to all polymer monomers, while solvent molecules feels indirect drag via monomers. In case of an oscillating field, a square-wave periodic potential is applied to each monomer given as $F_{G}^{x} =\text{sign}[\sin(2\pi \nu t)] G $. The  force is directed  along the channel axis,   where $\nu$ (time period, $\tau_\nu=1/\nu$)  stands for the frequency.

The physical parameters are expressed in units of  bond length $l_0$, energy  $k_B T$, and time  $\tau=\sqrt{m_s l_0^2/k_BT}$, where $m_s$ is mass of a solvent particle i.e., taken to be unity. The MPC parameters are collision time $\tau_s=0.1 \tau$, cell length $a=l_0$ and solvent density $\rho_s=10m_s/l_0^3$. 
Newton's equations of motion of the polymer is implemented using velocity-Verlet algorithm at fixed integration time step $h_m=5\times10^{-3}\tau$. All results are for chain length $N_m=200$, if not mentioned otherwise. We use a certain  mapping to obtain the experimental scales, equivalent to our coarse-grained model. This protocol has been previously found efficient in explaining many DNA based phenomenons  \cite{ali2006polymer,locker2006model}.A DNA of  width $w\approx2 $nm, $l_p\approx 50 $nm and contour $L=16 \mu $m is equivalent to a chain of  bead diameter $\sigma=0.8$, $l_p=20$, and $N=6400$. Considering each base pair to be of mass $615$ Da, a monomeric unit of length $2$nm (12 bps) has a mass of 7380 Da which is equal to $M=10$, a monomer's mass in simulations units. Hence, our simulation time unit $t=\sqrt{m_sl_0^2/K_BT}=1$  translates as $t=4.3\times 10^{-6}$ secs in real units. So, the chosen window of frequencies $0.0001-0.02$ maps to $23 Hz-4600 Hz$ for a chain length equivalent to $N_m=200$.  




\section{Results}\label{fx}

We present the structural response of a confined semi-flexible chain under an uniform  field. For that, we follow a systemic approach of addressing the influence of direct fields followed by an investigation into oscillatory fields.\\

\textbf{A. Chain collapse under direct field}\\

 \begin{figure}[t]	
\includegraphics[width=1\linewidth]{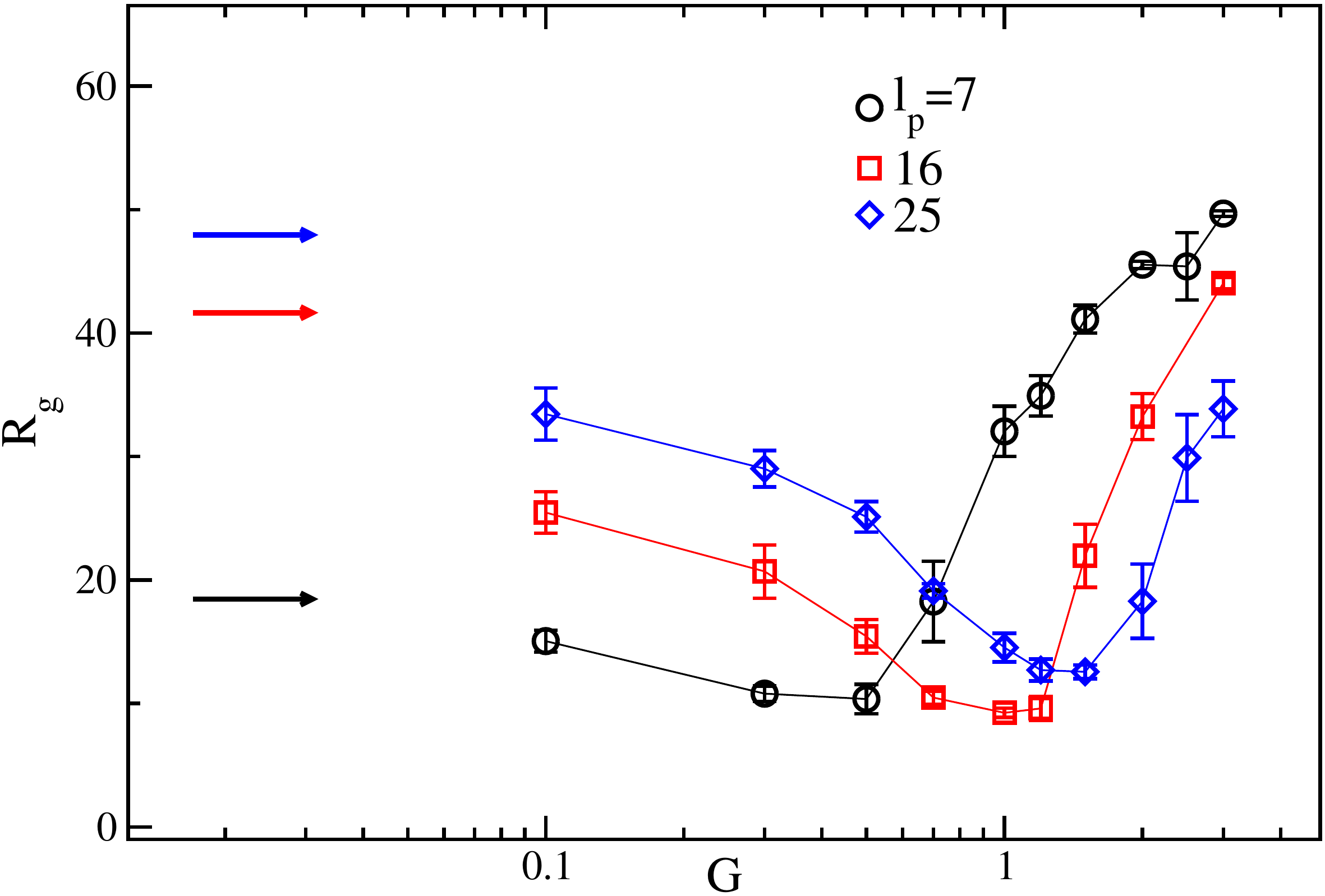}\\

 \caption{ a. Radius of gyration $R_g$  of a semi-flexible polymer under DC field $(G)$ for various persistence length  $l_p$ in pore radius $R_p=10$.     }
\label{fig:DC_vary_G}
\end{figure} 

 
 In  bulk, it is prestablished that  a short flexible chain  exhibits  weak compression under direct field\cite{Netz_JPCB_2003,Singh_2018_JCP}. 
 Figure~\ref{fig:DC_vary_G} elucidates the structural response of a confined semiflexible polymer ($N_m=200$) under subjection to a constant DC field. This is parametrized in terms of average radius of gyration, given as $R_g=\sqrt{\langle \frac{1}{N_m}\sum_{i=1}^{N_m}( {\mathbf r_i}-{\mathbf R}_{cm})^2\rangle}$, where ${\mathbf R}_{cm}$ is the center-of-mass of the chain. The retrieved curve exhibits a non-monotonic dependence over $G$, where within a moderate field strength chain  exhibits a significant structural compression, followed by a stretching  at higher field strengths\cite{Singh_2018_JCP}. This is qualitatively similar to the stretching response reported for the flexible chain in bulk, with the exception that there chain compaction was seen only for smaller chain lengths and weak field\cite{schlagberger2008anomalous,Netz_PRL_2009}. However, for a confined chain, not only the compaction  becomes prominent for $N_m=200$, the favourability of the induced compression shifts toward higher fields with increasing persistence length. Also, the relative compaction gets more pronounced for higher bending rigidities such as, for $l_p=25$ it is roughly $R_g/R_g^0=0.25$, while for $l_p=7$, $R_g/R_g^0=0.5$,  with $R_g^0$ being the radius of gyration of the chain in equilibrium. This is particularly fascinating considering the long and semiflexible nature of a typical DNA molecule\cite{hagerman1988flexibility}, which reportedly exhibits collapse under applied fields\cite{zhou_2015_collapse,Zhou_PRL_2011,Jing_PNAS_2011}. 
%

 \textbf{\textit{ Effect of bending rigidity:}}
 \begin{figure}

\includegraphics[width=\linewidth]{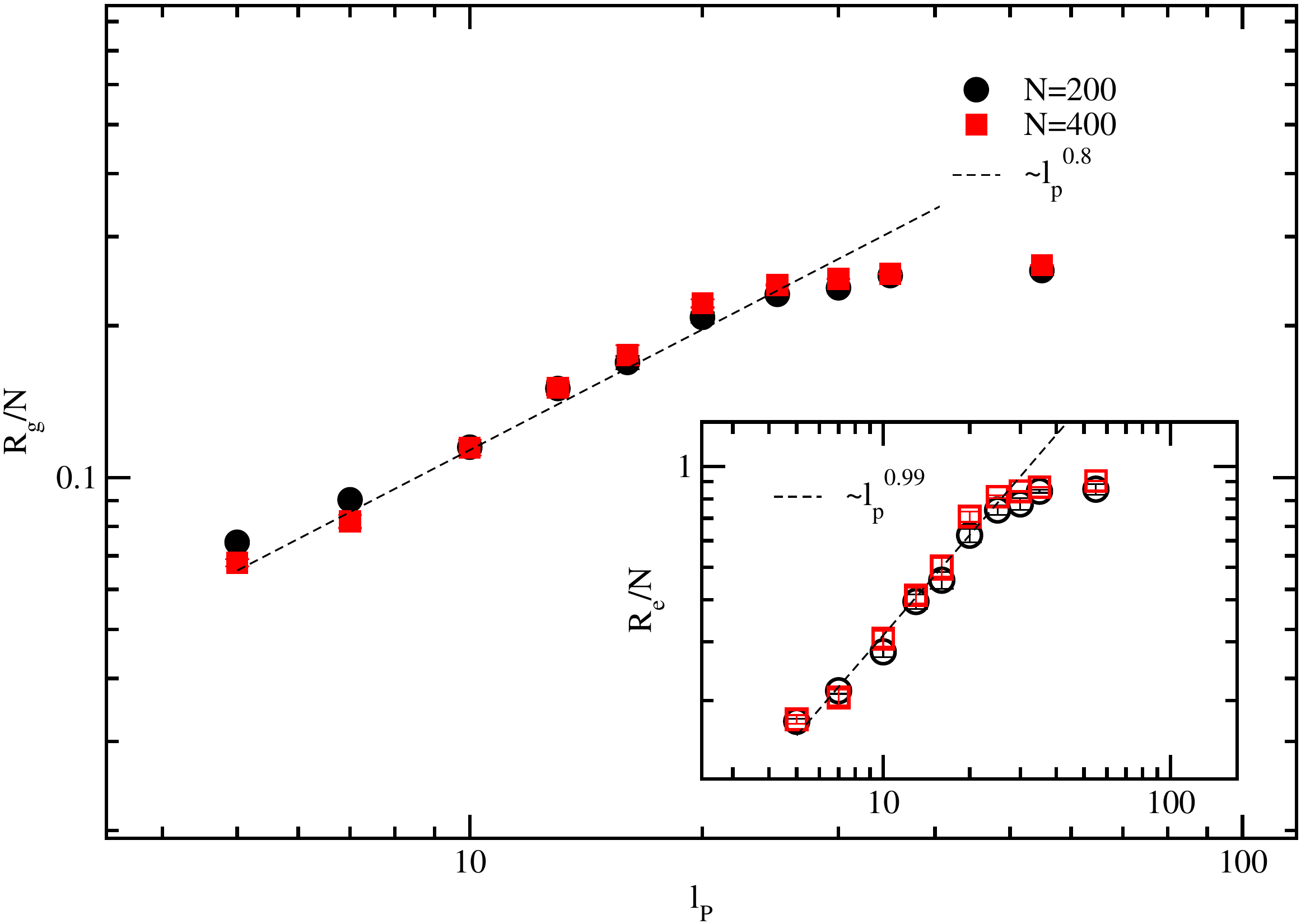} \\
\includegraphics[width=\linewidth]{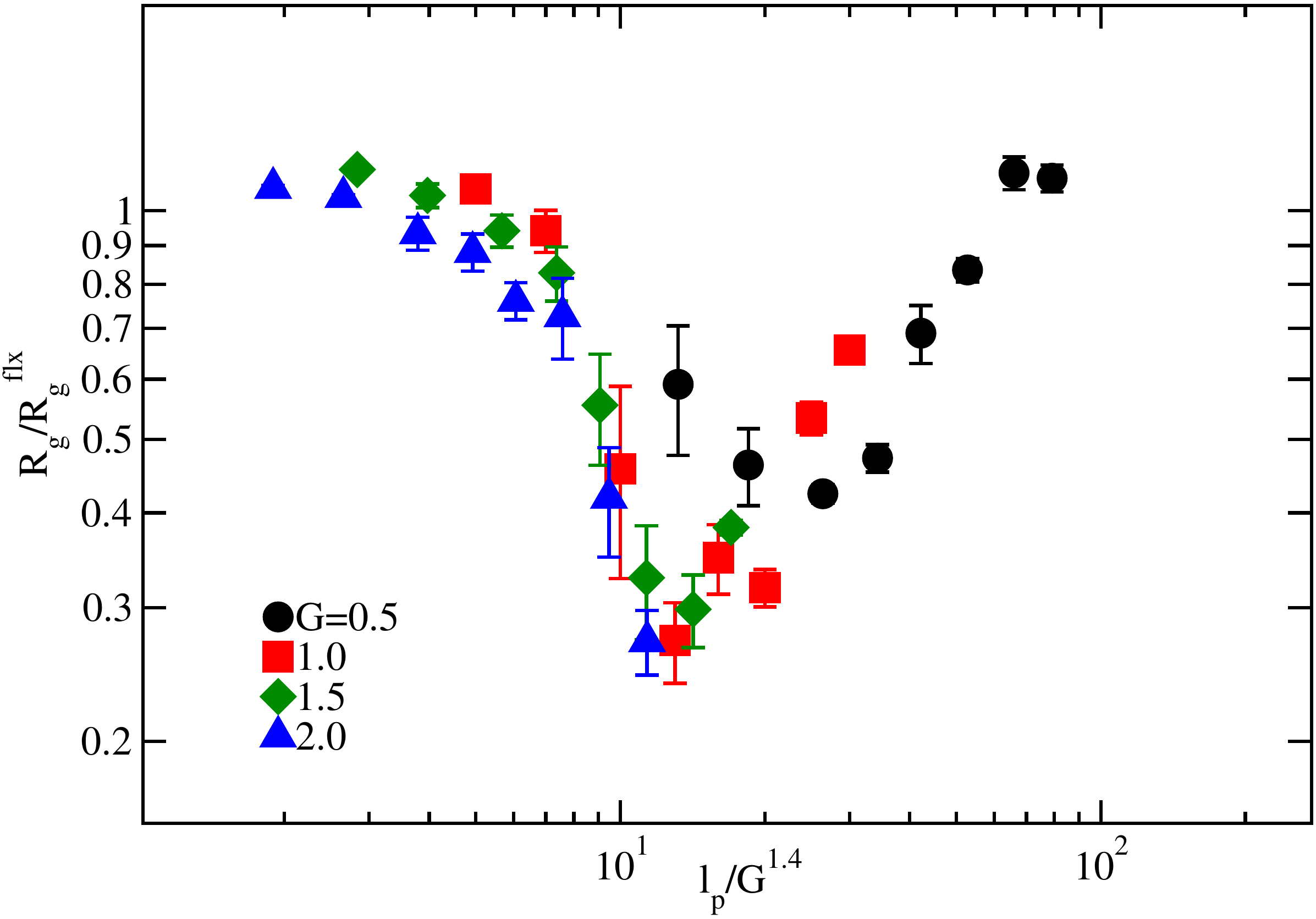} \\
\includegraphics[width=1.0\linewidth]{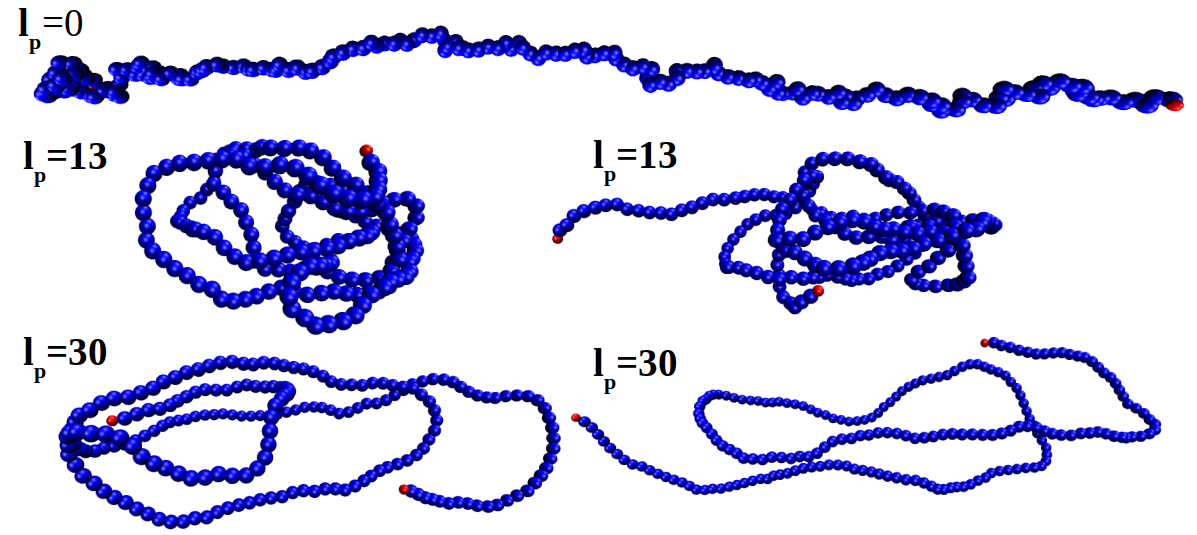}
\caption{ a) Scaled radius of gyration $R_g$  of a semi-flexible polymer chain with varying bending rigidity $l_p$ in confinement radius $R_p=10$ for two different chain lengths. The inset shows variation in end-to-end distance $R_e$. b) The $R_g$  of a confined chain as a function of $l_p$  for a fixed $G$ is shown in the main plot. The chain undergoes a collapse in DC beyond a critical $l_p$. c. Snapshots of the chain under DC for various $l_p$ at $G=1$ and $N_m=200$.} 
 \label{fig:DC_vary_lp}
\end{figure}
 In the equilibrium case, it is prestablished that a confined chain exhibits a monotonic stretching with increasing persistence length, where the chain with lower bending rigidity spanning the de Gennes regime/moderately confined de Gennes regime \cite{daoud1977statistics,wang2011simulation,reisner2012dna} ($D/l_p>>1$) takes a coil-like structure. However, with enhancing chain rigidity it transitions into Odijk's regime \cite{odijk1983statistics,odijk2008_PRE_scaling} ($D/l_p<1$), exhibiting stretching into linearly arranged array of the chain segments. The equilibrium curve is shown in Fig.\ref{fig:DC_vary_lp}-a .

  However, the chain exhibits a counter intuitive structural response with bending rigidity under direct field, which is shown in Fig.\ref{fig:DC_vary_lp}-b. The $R_{g}$ is normalized with the stretching  seen for the case of a flexible chain $R_{g}^{flx}$. Under DC, the monomers in the periphery experiences  an enhanced hydrodynamic drag compared to the monomers inside which drifts faster under field in a random coil-like chain\cite{schlagberger2008anomalous,Netz_PRL_2009}.  As a result of this spatial gradient in  drift velocity of monomers, a re-circulation flow field builds up that pushes shear lagged particles from the surface into the middle in a circulatory fashion. As a result for a flexible chain, this recirculating hydrodynamic flow field results in a tadpole-like structure\cite{Netz_PRL_2009,  schlagberger2008anomalous}, made of a compact head followed by an extended tail as shown in Fig.\ref{fig:DC_vary_lp}-c . 
 However, with increasing $l_p$, the size of this head grows, due to bending cost, leaving a folded structure constituting major fraction of the chain with shorter fluctuating tail. This manifests as the compression seen in intermediate $l_p$ regime, elucidated for $l_p=13$ in Fig.\ref{fig:DC_vary_lp}-c. However, beyond a certain $l_p$ the chain once again stretches (see $l_p=30$ in Fig.\ref{fig:DC_vary_lp}-c). At larger $l_p$'s, the bending stiffness dominates and disfavours large curvatures, giving rise to elongated backfolded domains.  In summary, the non-monotonic response  of the chain  under DC field is due to the competition between hydrodynamic drag induced compaction under field\cite{Netz_PRL_2009,  schlagberger2008anomalous}, and the bending energy cost.

 

 A similar non-monotonicity 
 is also observed in the spontaneous knot formations seen in polymers\cite{zhu2021revisiting,coronel2017non}. This similarity can be drawn, considering  the self-entangled and knotted structures seen in our case, which will be corroborated in later parts of the manuscript.  A heuristic scaling obtained for the critical bending rigidity, beyond which the chain starts compressing indicates $l_p^c \sim  G^{\beta}$ kind of dependence, where $\beta\approx 1.4$. 
 
  \textbf{\textit{ Effect of confinement:}}
  \begin{figure}	
\includegraphics[width=1\linewidth]{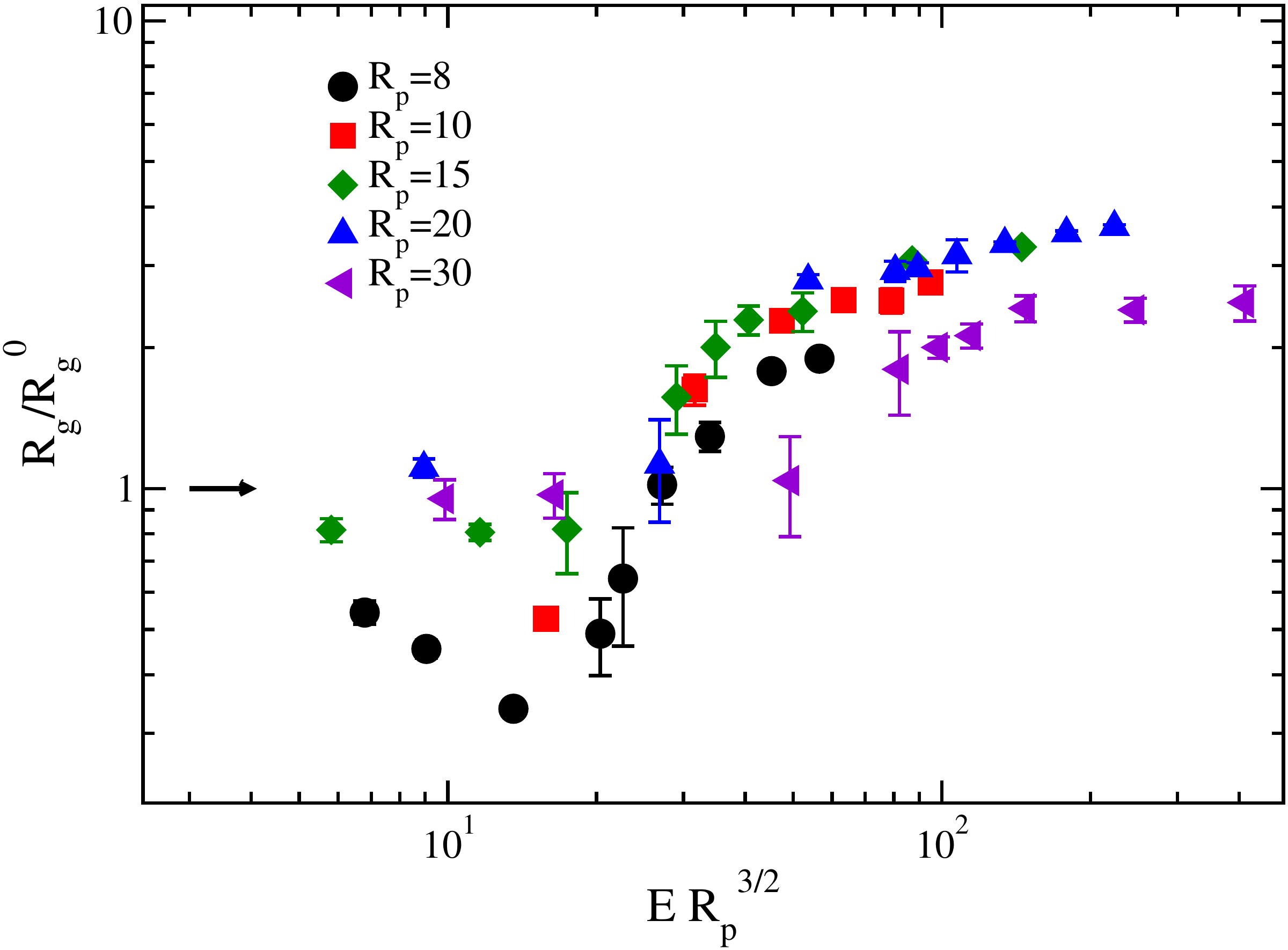}\\
\includegraphics[width=\linewidth]{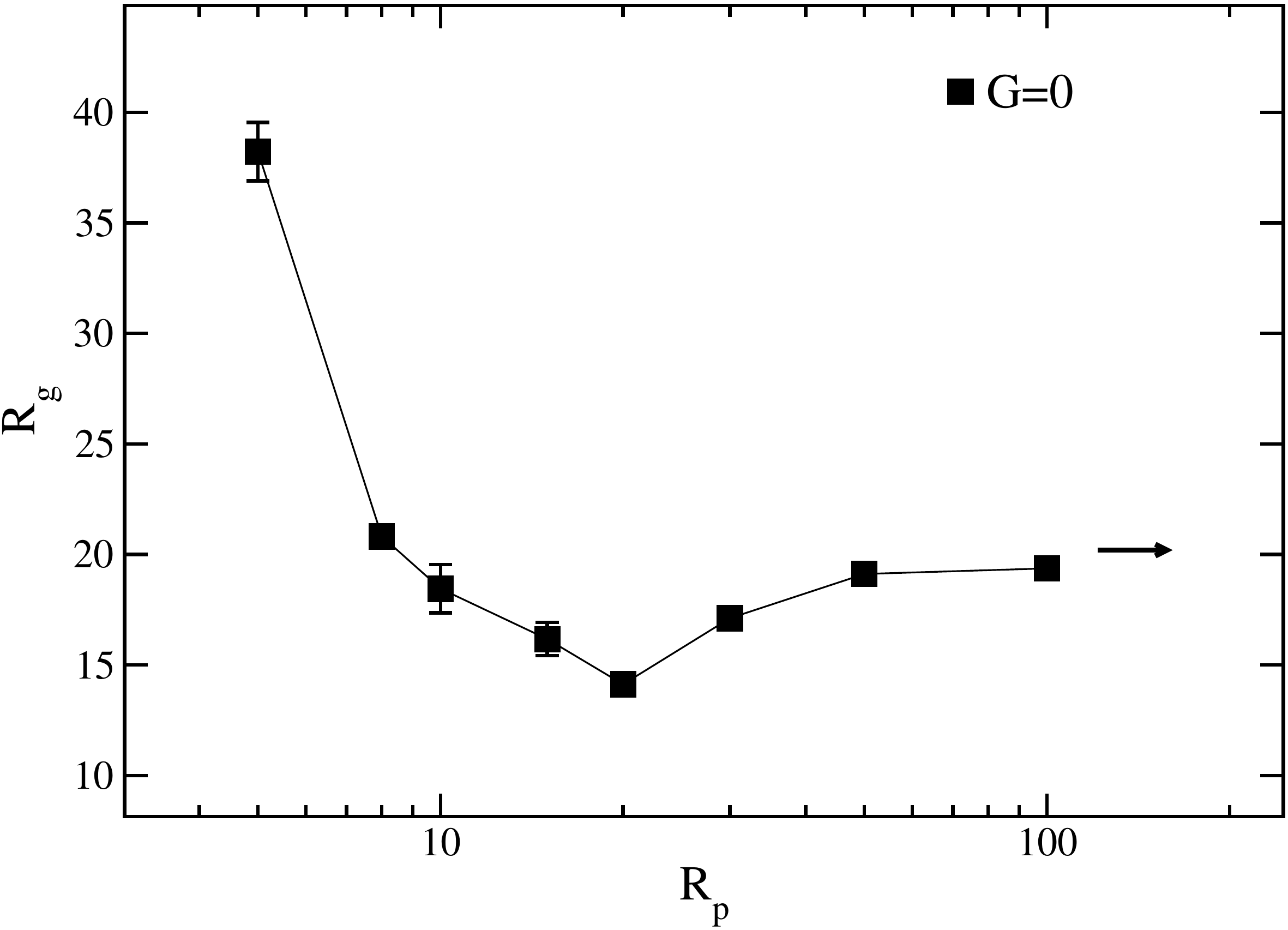}

 \caption{  a. The normalized radius of gyration of  a chain at $l_p=7$, as a function of $G$ for a range of confinements. Note that the x-axis is scaled with  $R_p^{3/2}$. b. The equilibrium radius of gyration of a confined semiflexible chain as function of pore radius at $l_p=7$ for chain length $N_m=200$.
 }
 \label{fig:dc_results}
\end{figure}
 Considering how the presence of confinement affects the conformational dynamics of a chain\cite{odijk2008_PRE_scaling,reisner2012dna,wang2011simulation}, its outright indispensable to look at the influence of the geometric constraint. Figure~\ref{fig:dc_results}-a  elucidates the field induced compression of  a semi-flexible polymer ($l_p=7$), for a range of varying pore radii. A striking observation is that, while moderately confined polymer, $D/2l_p < 1.4$,  undergoes a substantial compression at lower field strengths followed by a stretching, chain under weak confinement/tending to bulk $D/2l_p>1.4$ exhibits no compression and undergoes a monotonic stretching. This is evident in $R_p \leq 10 $ where a compressive dip in $R_g$ is seen for lower G values, while for $R_p>10$ only monotonic stretching is seen, devoid of any shrinkage. A heuristic scaling obtained in Fig.~\ref{fig:dc_results}-a for the field $G_c$ beyond which stretching is effectuated, gives a $G_c\sim R_p^{-3/2}$ dependence. 
 The deviation in $R_p$ scaling seen for $R_p= 30$, we speculate is a repercussion of $R_p$ hitting a different regime, where the effect of confinement diminishes such that all $R_p > 20 $ tend to the bulk behaviour. This is elucidated in Fig.\ref{fig:dc_results}-b, which shows the structural variation of a chain under varying confinement in equilibrium. It shows a typical non-monotonous pattern, as reported earlier\cite{cifrasimu2012}. The chain slightly swells back to its bulk value beyond $R_p>20$.  \\

 \begin{figure}
\includegraphics[width=\linewidth]{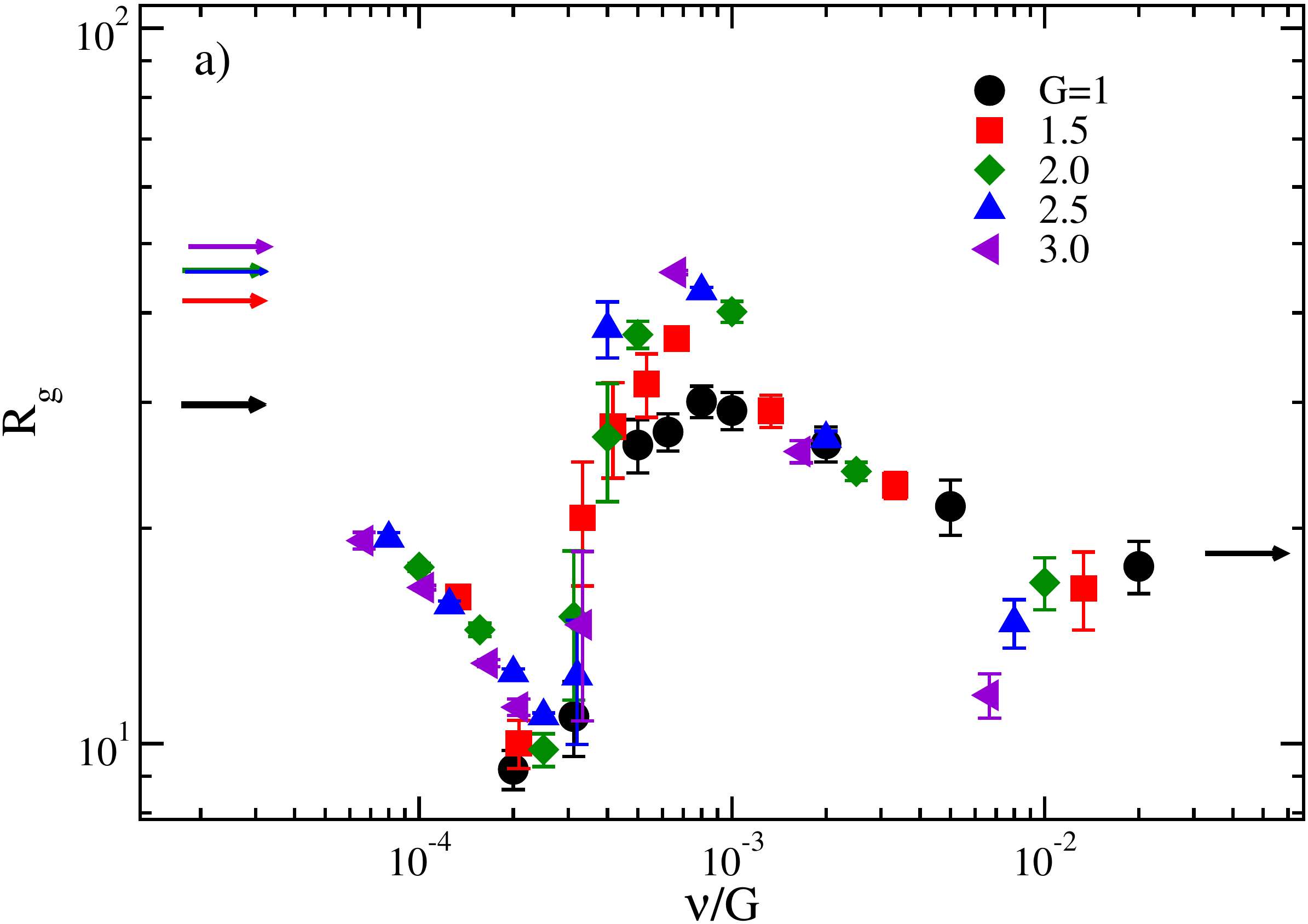}
\includegraphics[width=1.0\linewidth]{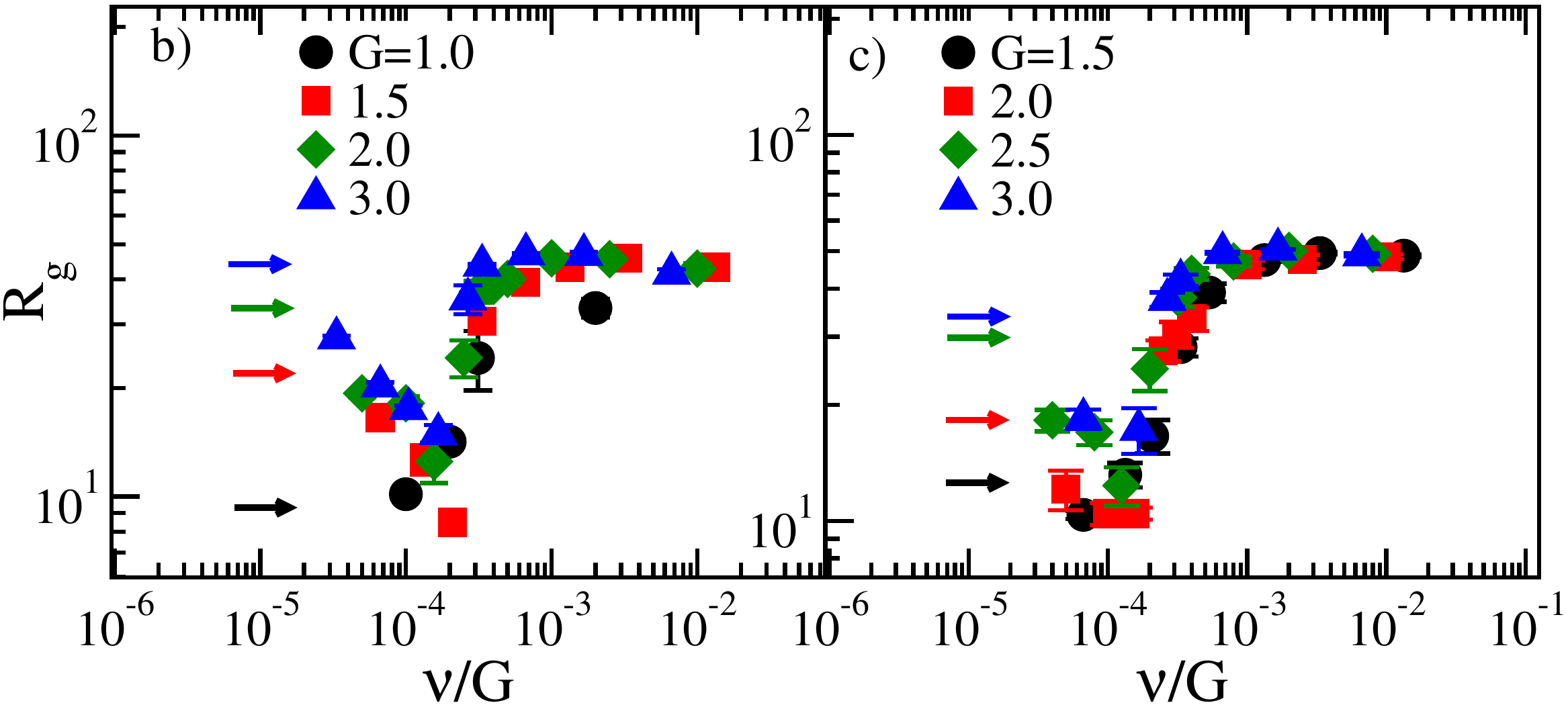}
\caption{ The  variation of $R_g$  in response to varying field frequencies for different  $G$ and fixed $R_p=10$. Different panels correspond to  persistence length a) $ l_p=7$, b) $ l_p=16$, and c) $ l_p=25$, respectively. The $R_g$ for the respective DC values  are denoted by a horizontal arrow. }
\label{fig:AC_varylp}
\end{figure}





\textbf{B. Chain under Oscillating-Field}\\

\begin{figure}	
\includegraphics[width=\linewidth]{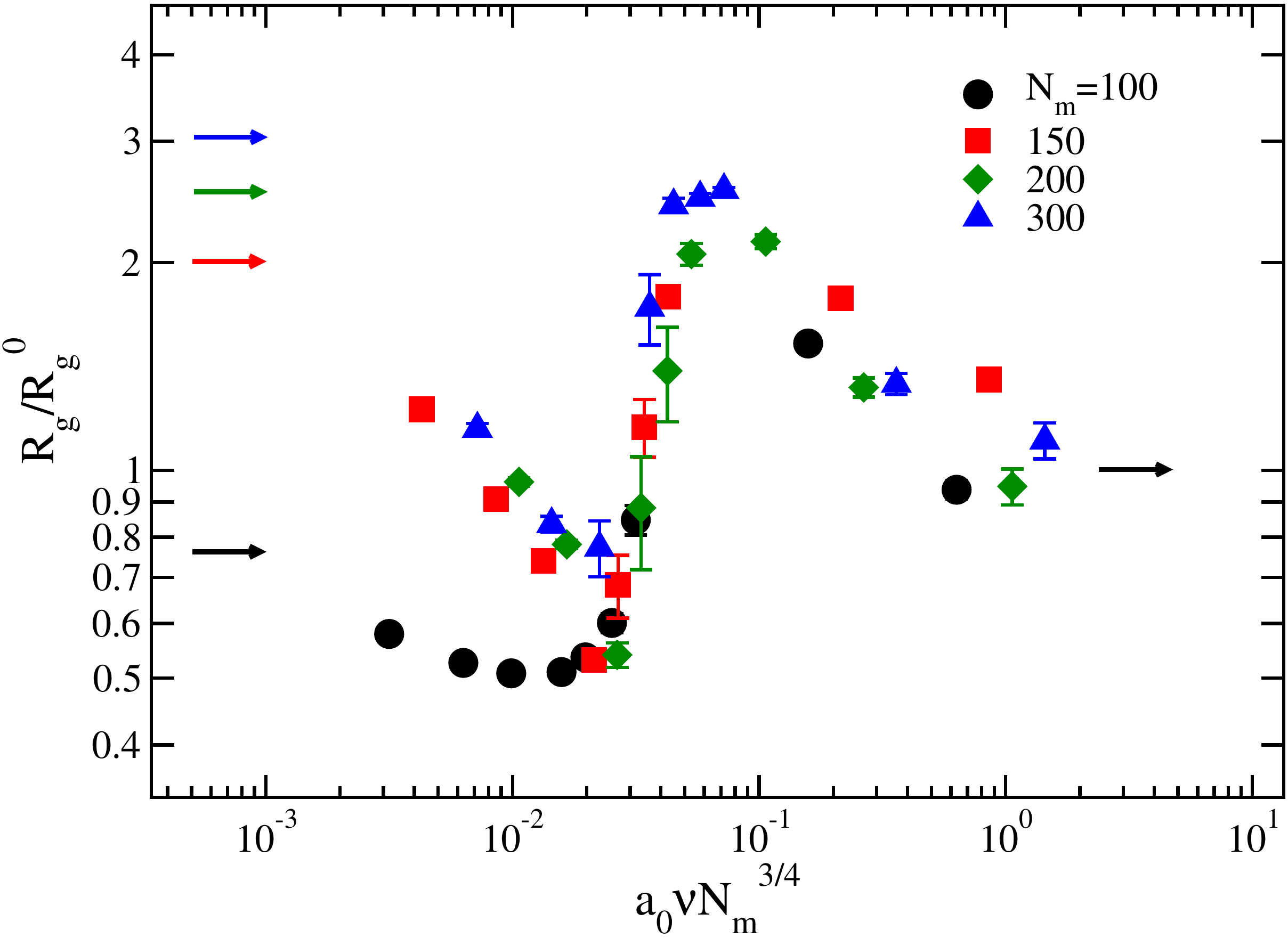}\\
   \caption{{ Scaled radius of gyration $R_g$  of polymer chain in pore $R_p=10$ in response to varying field frequencies at $G=2$, for different chain length $N_m$. Horizontal lines denote the $R_g$ for the same DC field. }}
    \label{fig:vary_N}
\end{figure}

The structural response of the chain under an oscillating field  is shown in Fig.~\ref{fig:AC_varylp}   for  $l_p=7$ and   $R_p=10$. At higher frequencies the chain remains nearly unperturbed retaining the equilibrium structure ($R_g^0$).  
In the intermediate frequencies, the chain  swells  along the field direction, referred as the "\textit{stretch-state}",  until it reaches a critical frequency $\nu_c$  below which the chain undergoes a significant collapse. This "\textit{compressed-state}" essentially consists of a crumbled folded domain with a short fluctuating tail.  
This collapse is further followed by a stretching at lower frequencies approaching toward the DC like behavior. Unlike the stretch-state mentioned earlier, the structure of the chain is more like a "\textit{tadpole}" with a leading bob-head and trailing long tail, that switches the  direction at every field switch. This eventually merges with the DC limit, where for long chains a tadpole-structure is found  with a long trailing end\cite{schlagberger2008anomalous,Netz_PRL_2009,Singh_2018_JCP}. In summary, for a range of $\nu<<\nu_c$, $\nu<\nu_c$, and $\nu \geq\nu_c$, the chain undergoes a three state transition involving \textit{tadpole-collapse-stretch} states, respectively (see SI-movie-1). 
 Additionally, we see similar behavior for various chain lengths  ($N_m=100,150,200,$ and $300$), where the chain compression for longer chains happens at lower frequencies. This dependence over chain length is shown in Fig.\ref{fig:vary_N}. Further, a  scaling of $\nu_c \sim N_m^{-3/4}$ for critical frequency with chain length  is retrieved by superimposing all the curves on to each other.  

 A scaling of $R_g$ with $\nu/G$ is procured, where all the curves superimpose on each other, suggesting a linear dependence of $\nu_c$ with G. The striking observation here is that for $l_p=7$,  the chain stretches under DC field for $G>1$ (see of Fig.~\ref{fig:DC_vary_G}). However, the oscillating field drives the system to a compressed state for the same field strengths. Similar AC induced compressions are seen for other bending rigidities  $l_p=16$ and $l_p=25$ (see SI-movie-2), shown in Fig.~\ref{fig:AC_varylp}-b, and c for a wide range of  $G$ beyond linear response regime, which is devoid of chain foldings in DC. Another, impressive feature of this AC driven compression is that same extent of collapse is attainable across all these field strengths. While, the maximum collapse possible varies from $R_g/R_g^0=0.55, 0.25, 0.2$ for  persistence lengths $l_p=10,16,$ and $25$, respectively.   Further, the transition points for the collapse obtained in  Fig.~\ref{fig:AC_varylp}-a, b, c suggests that the retrieved critical frequency $ \nu_c/G $ is nearly independent of the persistence length of the chain.
 
\begin{figure}[t]	
\includegraphics[width=\linewidth]{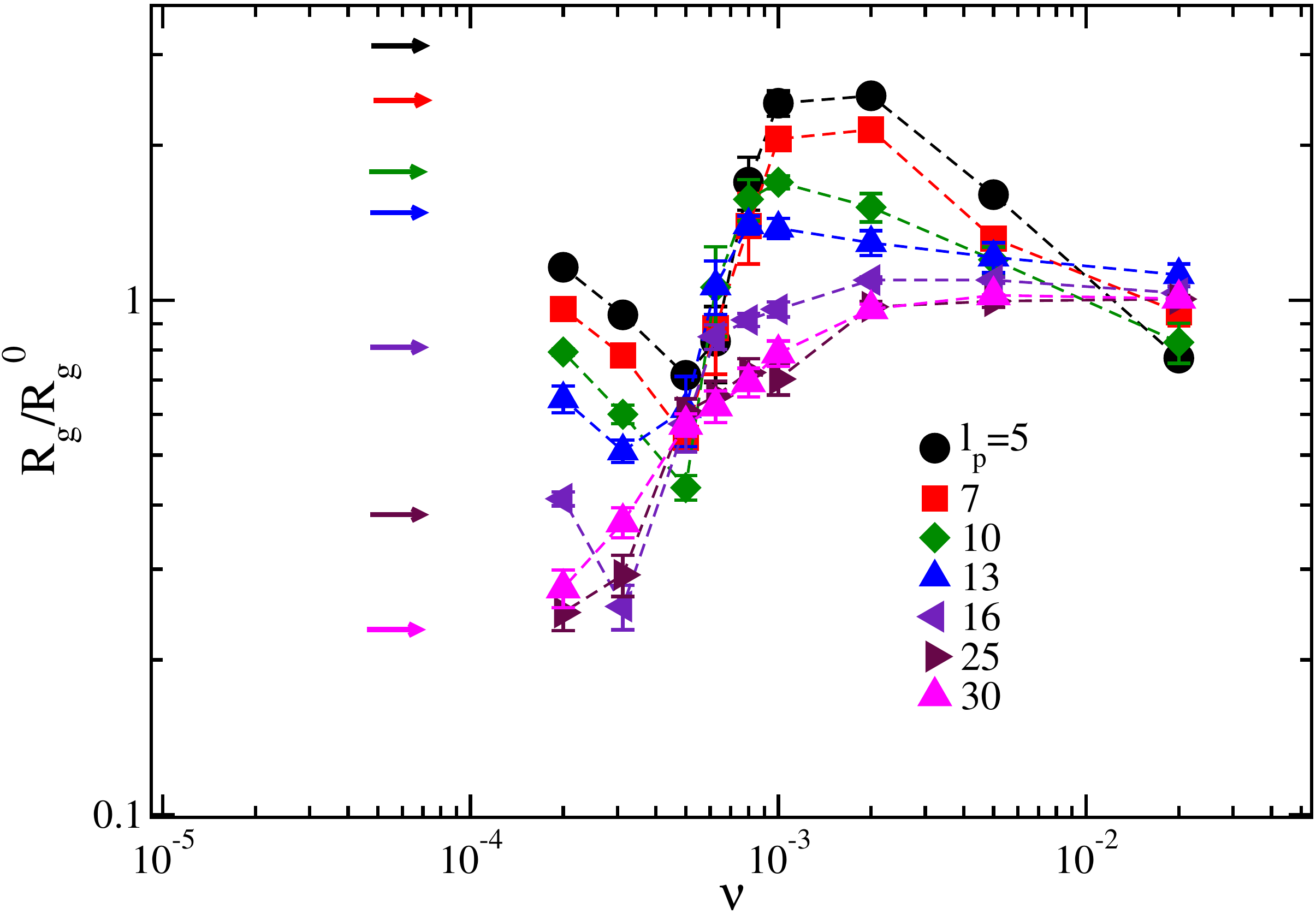}\\
   \caption{{ The normalized radius of gyration  $R_g/R_g^0$  of the chain at $R_p=10$ in response to varying frequency at $G=2$, for different persistence length $l_p$. The horizontal arrows denote the DC values of $R_g/R_g^0$ ($\nu=0$). }}
\label{fig:vary_lp}
\end{figure}

 \textbf{ \textit{Effect of bending rigidity under AC field:} }
 Now we consider the explicit effect of bending rigidity on the chain's collapse mechanism.  This is shown in Fig.~\ref{fig:vary_lp} for different bending rigidities at $G=2$ and pore radius $R_p=10$. More flexible chains undergo a {\it stretch-collapse-tadpole} state transition, while going from higher to lower frequency in the respective window of $\nu \geq\nu_c$, $\nu<\nu_c$, and $\nu<< \nu_c$. The stretching observed at higher frequencies, prior to the compression  is suppressed with increasing  bending stiffness. This again indicates toward the favoured chain stretching along the field direction, due to enhanced geometric constriction. For example, beyond $l_p>13$ stretching is completely absent, which corresponds to $D/2l_p^c \geq 1.0$, that is close to the ratio obtained in the case of the pore radius variation. The discrepancy in the crossover regime here, we speculate is due to  shorter chain lengths ($N_m=200$) used in the simulations. In summary, for a confining radius, there exists a critical persistence length $D/2l_p^c \approx 1.0$, beyond which the chain eludes the pre-stretching tendency before collapse  at high frequencies ($\nu>\nu_c$). Also, for higher persistence lengths, the folding favourability enhances dramatically and the chain exhibits collapse for a wide range of frequencies, even extending up to the DC limit. 

\begin{figure}[htb!]
\includegraphics[width=\linewidth]{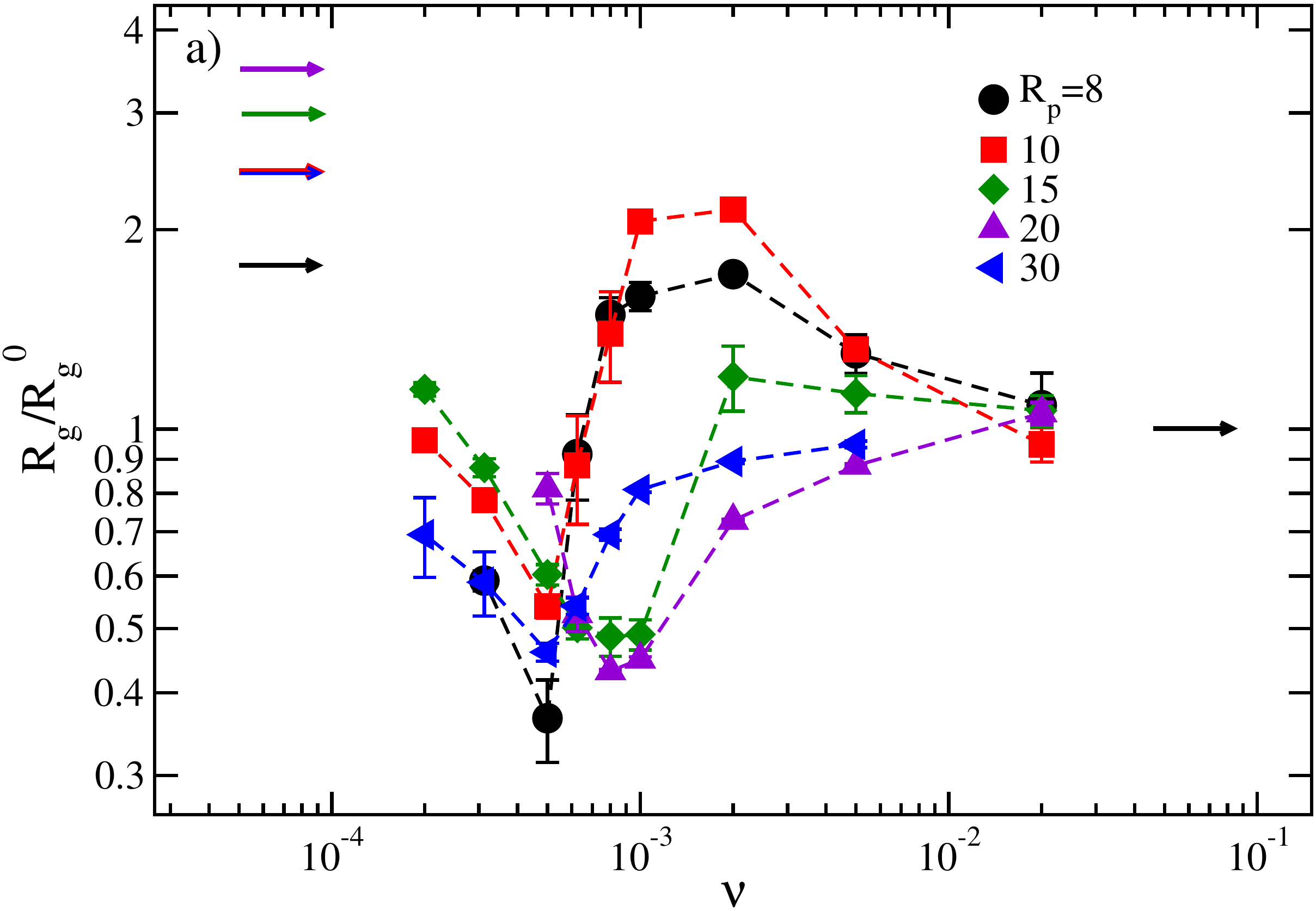}
\includegraphics[width=\linewidth]{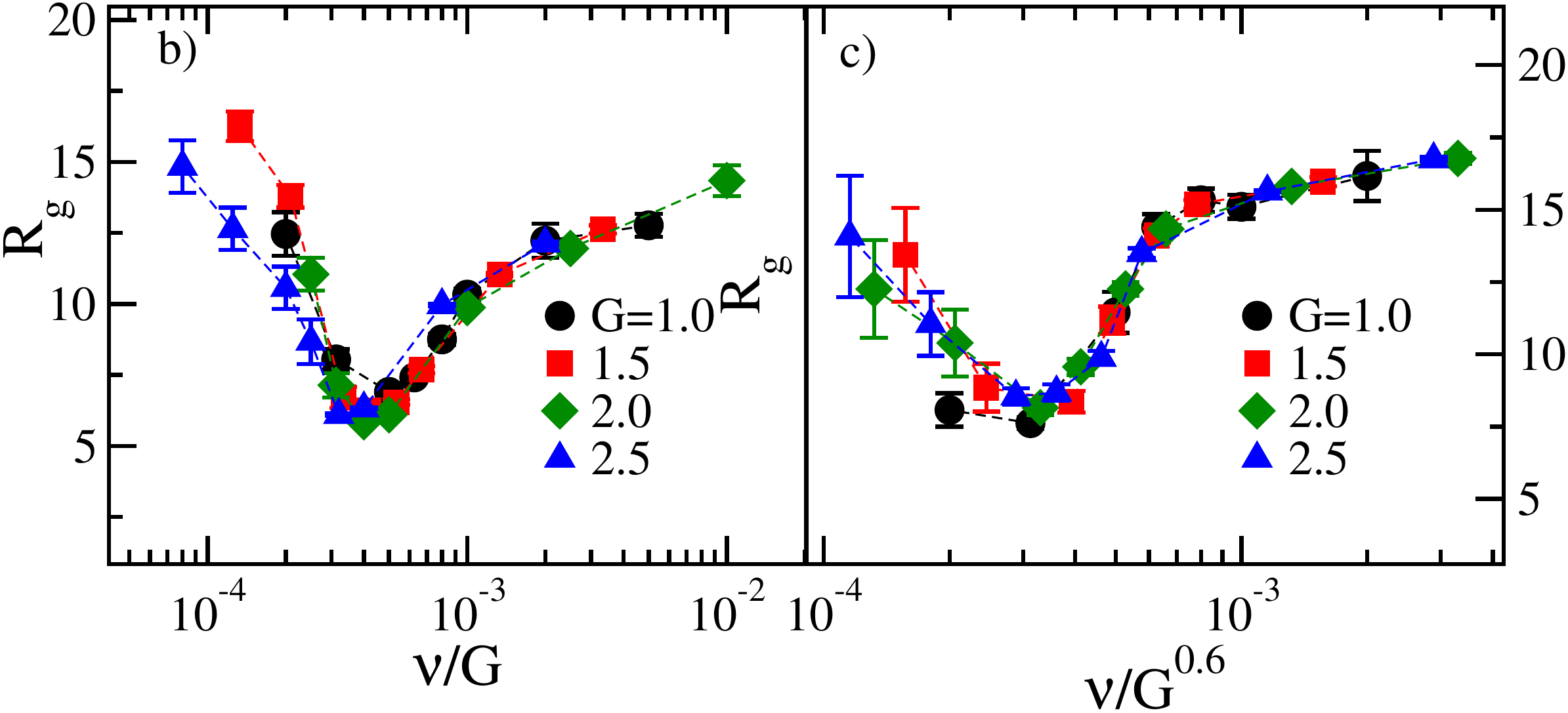}
   \caption{ (a) Scaled radius of gyration $R_g$  of semi-flexible chain in confinement in response to varying field frequencies at $G=2$, for different pore radius $R_p$. Horizontal lines denote the DC values.  Here, (b)  and (c) corresponds to the variation in $R_g$  of polymer chain in confinement in response to frequencies for different field strengths G, for pore radii $R_p= 20$ and $30 $, respectively.   }
    \label{fig:vary_pore}
\end{figure}
\textbf{\textit{ Effect of confinement under AC field:}}
The effect of confinement on the structural response of a chain for different pore radius  is displayed in Fig.~\ref{fig:vary_pore}-a,  at $l_p=7$ and $G=2.0$. For a narrow pore $R_p\leq10$, with decreasing frequency a three tier stretch-collapse-tadpole kind of transition is observed.  While with increasing pore radii, the stretching prior to collapse (for $\nu\geq\nu_c$) diminishes, as is evident in the case of $R_p=15, 20$, and $30$, which is devoid of such stretching. 
\begin{figure}	
\includegraphics[width=\linewidth]{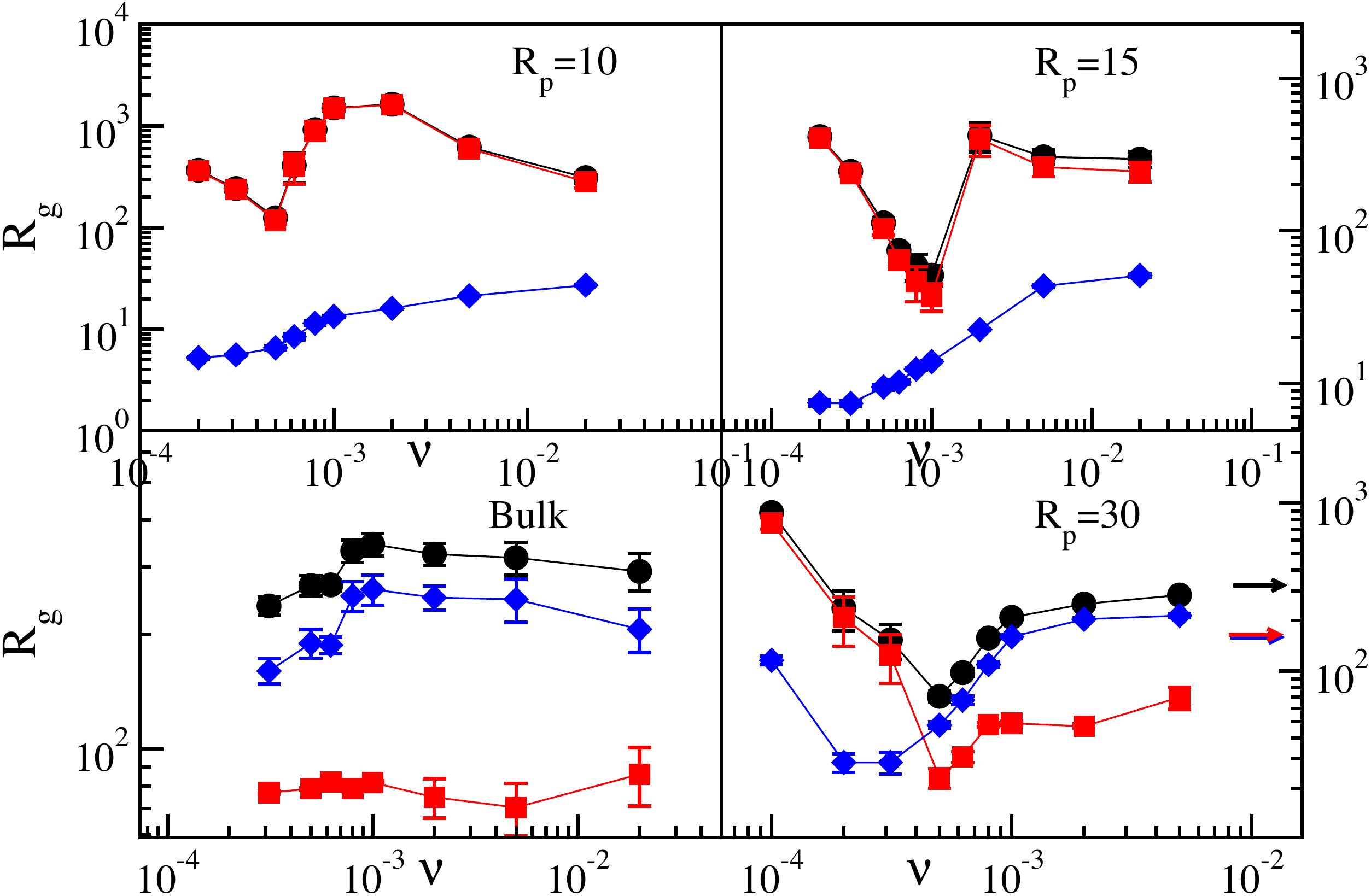}\\
   \caption{Transverse and longitudinal components  of radius of gyrations $G_{xx}$ and $G_{yy}+G_{zz}$ for  a few  pore radius, including bulk. The values  correspond to $N_m=200$ and $l_p=7$ for all pores, while bulk is shown for $N_m=50$, $l_p=7$).  
   The stretching in lower $R_p=10$ is effectuated by pore enhanced elongation along the field-axis. The stretching in bulk is attributed to the stretching and alignment of the chain along the transverse direction.}
    \label{fig:rg_comp}
\end{figure}

To gain insights into various  stretching responses, we distinctly look into the chain's expanse in the longitudinal and transverse directions to the applied field. Figure~\ref{fig:rg_comp} shows the radius of gyration $R_g^2={\langle \frac{1}{2N^2}\sum_{i,j}( {\mathbf r_i}-{\mathbf r_j})^2\rangle}$ (in black circles), its component along x-direction of confinement axis $G_{xx}={\langle \frac{1}{2N^2}\sum_{i,j}( {x_i}-{x_j})^2\rangle}$ (in red squares) and its component in the perpendicular y-z plane $G_{yy}+G_{zz}={\langle \frac{1}{2N^2}(\sum_{i,j}( {y_i}-{y_j})^2+\sum_{i,j}( {z_i}-{z_j})^2)\rangle}$ (in blue diamonds) for varying pore radii $R_p=10,15,30$, and bulk depicted in clockwise manner from top left, respectively. Interestingly for $R_p=10$, the $G_{xx}$ curve maps the overall $R_g^2$, suggesting that the stretching seen here is by virtue of the chain extension along the channel axis. With the transverse component contributing barely, this gives a highly anisotropic stretched state. However, with increasing pore radii, for $R_p=15$, the x-component departs away from the overall $R_g$ and the contribution from the transverse component enhances, decreasing the anisotropy along the field direction. Interestingly, with  further increase in pore radius like for $R_p=30$ the structure again shows an increasing anisotropy, by preferentially aligning in the y-z plane. Further, tending to the bulk case the chain again exhibits prominent transverse alignment for higher frequencies, where $(G_{yy}+G_{zz})/G_{xx} > 2$. Hence, the stretching seen for $R_p=10$ and below in Fig.~\ref{fig:vary_pore}-a is a consequence of stretching along the field direction with transverse alignment of the chain constricted by the narrow wall, in contrast to the bulk where the chain preferably stretches and orients in the transverse direction. Also, it can be inferred that the chain undergoes a direct collapse, bypassing the stretched state at higher frequencies, approximately in  the regime $D/2l_p>1.4$.

 Additionally, a non- monotonic shift in  the critical frequency $\nu_c$ with pore radius is  seen, where beyond  $R_p=20$ the $\nu_c$ decreases.  This maybe stemming from similar non-monotonicity in $R_g$ seen in equilibrium fig.\ref{fig:dc_results}-b, which reflects in the associated timescales as well. 






Further, structural response of the chain obtained for $R_p=20$, and  $R_p=30$ for varying fields shown in Fig.~\ref{fig:vary_pore}-b,c, exhibits a strong coil to compressed-state transition at the critical frequency, before tending to the DC limit. See SI-movie-2 for compressed states. 
Here,  we retrieve a scaling for the critical frequency with field strength as  $\nu_c\sim G$ for $R_p=20$, similar to $R_p=10$ case, while for $R_p=30$ we get $\nu_c \sim G^{3/5}$. This difference is speculated to be a result of the chain hitting the bulk regime beyond $R_p=20$, see Fig.\ref{fig:dc_results}-b. Previously, we have seen that in DC for
narrow pores, strong compression is seen at lower fields in contrast to wider pores (like $R_p=20$ and $30$), where no such folding is seen (see Fig.~\ref{fig:dc_results}-a). Interestingly, under oscillating field not only narrow pores( $R_p=10$ in Fig.~\ref{fig:AC_varylp}-a) even larger pores ($R_p=20$ and $30$ in Fig.~\ref{fig:vary_pore}-b,c)  exhibit remarkable compression of the chain for a range of field strengths, where chain remains stretched under DC field. Hence, AC driven collapse mechanism is accessible over a wider range of  confining radii (including bulk) and field strengths beyond the linear response regime.   \\

\textbf{C. Transient dynamics under DC:}\\

\begin{figure}	
\includegraphics[width=\linewidth]{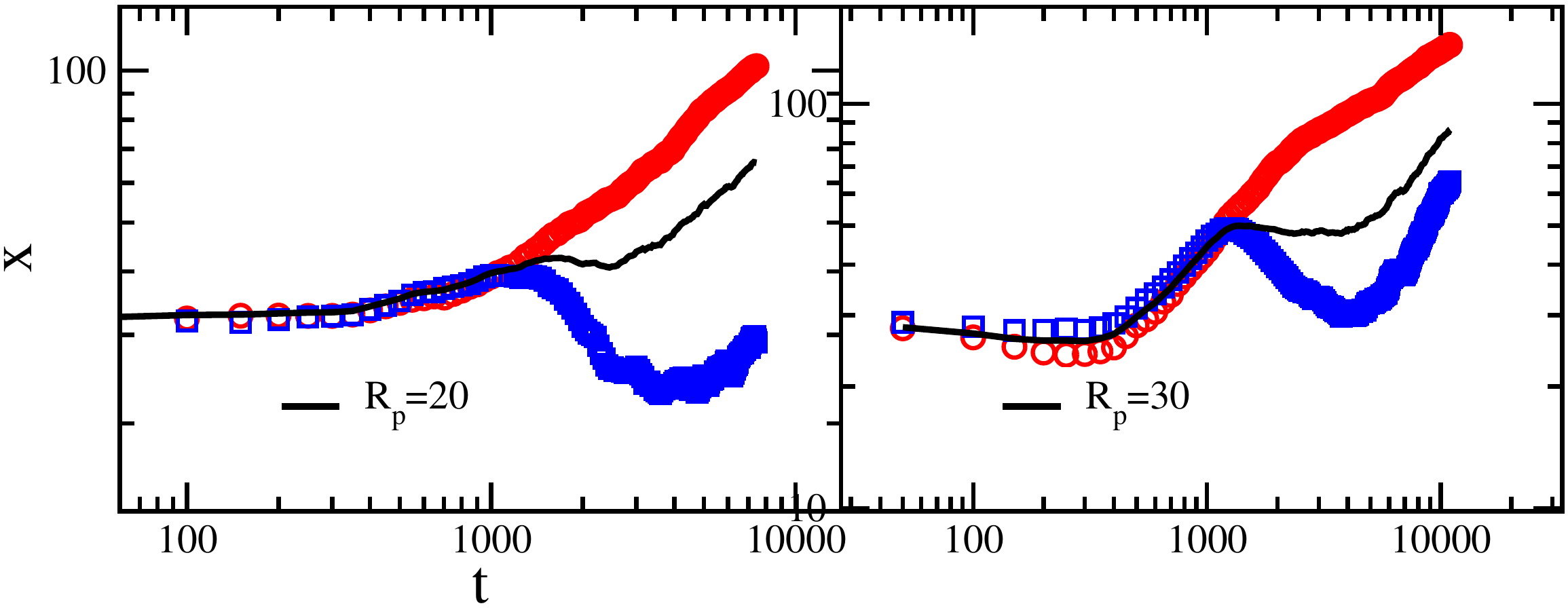}
  \caption{The average maximum extension of chain as a function time under DC field  for pore radius $R_p=20$ (Left panel) and $R_p=30$ (Right panel) at $G=2.0$.  The solid line shows the average over all the confirmations.  The red line shows the average over trajectories which undergoes straight stretching, while the blue is for those which undergoes an initial compaction followed by stretching.} 
    \label{fig:dynamics}
\end{figure}
  The dynamic pathway of the tadpolar formation under DC field involves an initial compaction followed by an extending tail\cite{Netz_PRL_2009,schlagberger2008anomalous}.  We corroborate this by conducting 50 independent trials of a chain under DC, where majority of conformations undergo an initial compaction into a compressed-state  prior to the stretching,  while a few others undergo direct stretching, where a small head like structure  grows over time to form an extended tadpolar structure.  The transient dynamics obtained categorically by averaging over respective events (with and without initial collapse) are shown in Fig.~\ref{fig:dynamics}-(a) and (b). Interestingly,  a chain which  started from the coil-state attains a collapsed-state, on average within the time window of $t_c=1000-10000$. This time-scale translates to frequency as $\nu=1/t_c=0.001-0.0001$, which coincides with the frequency window of chain collapse in AC (see Fig.~\ref{fig:vary_pore}-b and c).  The transient dynamics of the chain under homogeneous DC field  indicates that if for an oscillating field, the allowed time window before every field switch is such that it captures the initial collapse process, partially forbidding the complete tail extension, the chain wouldn’t relax to its actual DC field extension values (see Fig.~\ref{fig:dc_results}-a). 

\begin{figure}[t]
\includegraphics[width=\linewidth]{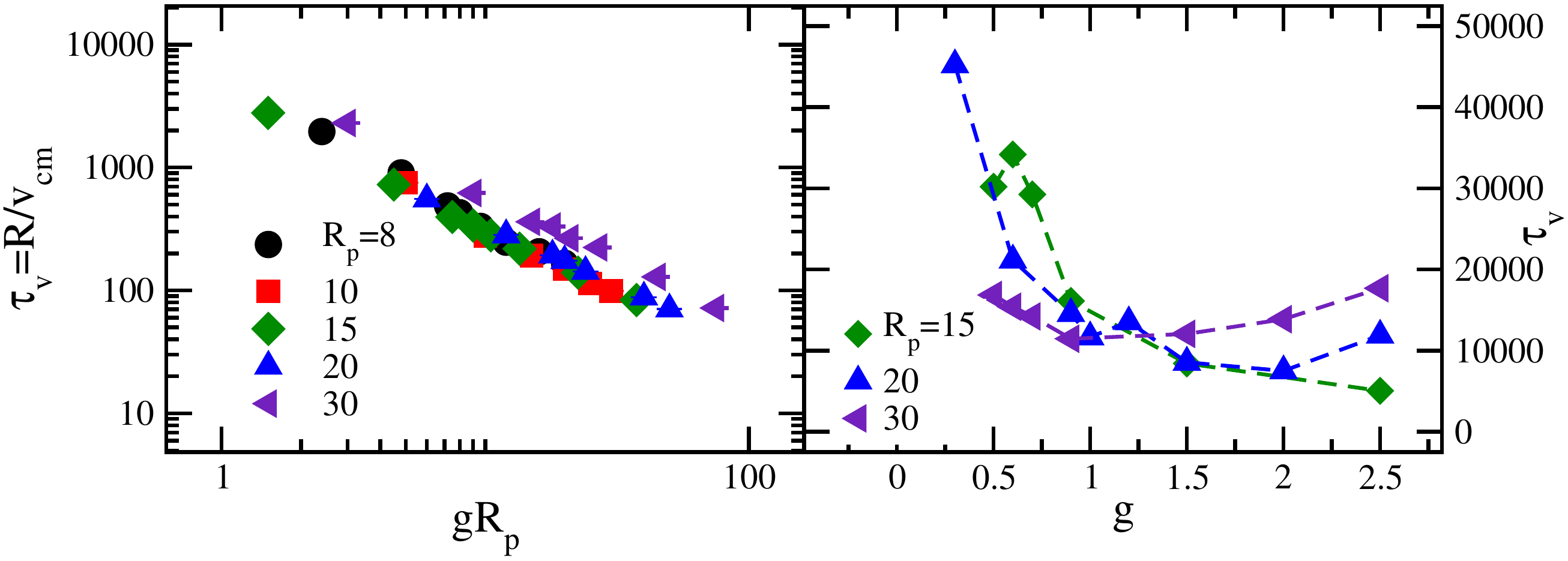}
  \caption{ a) Recirculation timescale $\tau_\nu=R_0/v_{cm}$ for different pores as a function varying field strength $G$ in DC. b) The  DC relaxation timescale obtained from the correlation of the max extension.  }
    \label{fig:cutoff_time}
\end{figure}

Further, it is prestablished that the chain compaction under DC, is a result of a recirculating flow field. Under such a scenario, a diffusive timescale of the chain can be given as   ($\tau_D=\frac{R^3}{\mu_0ak_BT}$), where the chain traverses a distance of its own size in equilibrium. Similarly, there exists a drift timescale of the chain ($\tau_v=R/v_G$), which essentially signifies the time any part of the chain takes to traverse a distance of R under the field, enforcing proper recirculatory dynamics. The later is a precursor to obtain chain compaction under DC.  Hence, in case of  an oscillating field at  $G$,  if the time-period of the applied field follows 
$1/\nu>\tau_v(G)$ (i.e. $\nu<1/\tau_v(G)$),  the chain is allowed proper recirculating dynamics leading to a compact state in AC. This recirculation timescales  as a function of  $G_{DC}$  is presented in  Fig.~\ref{fig:cutoff_time}-a. Since in our case, $\tau_v$ spans a window of $\tau_v=100-1000$, roughly, translating as $1/\tau_v=0.01-0.001$, then $1/\tau_v$ provides a frequency window  below which chain compaction can be obtained. However, at much lower frequencies (i.e larger allowed times) under field, the tail extension again gets prominent leading to the DC limit. As a result the compaction is seen only in a narrow window.  

While, this roughly sets the upper frequency cutoff for the collapse to reinforce,  
the lower frequency cutoff for the oscillating field below which the chain reaches to the DC limit, is obtained from the relaxation of the max extension under the DC field. We estimate the  relaxation of the maximum extension  of chain defined as  $C_X(t)=\frac {\langle \delta X(t) \delta X(0) \rangle}{\langle \delta X(0) \delta X(0)  \rangle}$, where $\delta X=X(t)-\langle X(t) \rangle$ is the fluctuation in maximum extension, and $X(t)$ and $<X(t)>$ are the instantaneous and average max extension, respectively. The correlation shows exponential decay in short time limit followed by oscillatory behavior for large time. The behavior of the correlation can be parametrized  by the following expression,  $C_X(t)=a_0 \exp(-t/\tau) \cos(\frac{2\pi t}{T})$.  The retrieved zero crossing time $\tau_X^{DC}$ at which $C_X(\tau_X^{DC})=0$ values are shown in Fig.~\ref{fig:cutoff_time}-b, for varying $G$ with different pore radii. The obtained relaxation time $\tau_X^{DC}$ falls in the range of $10000 - 40000$, which translates to  $1/\tau_X^{DC}=0.0001 - 0.000025$.  This signifies that beyond $\tau_X^{DC}$ the chain attains proper conformational relaxation of its fluctuating head and tail ends. This results in stretched tadpolar states seen in the frequency window $\nu<<\nu_c$, corresponding to the  DC limit.  \\

{ 


\textbf{D. Field induced knotting:}\\

Now our goal is to corroborate the presence of knots driven by the AC field and  how  frequency modulation influences the knotting tendency and its complexity.
For this purpose,  we  use a software package \textit{Kymoknot}\cite{tubiana2018kymoknot},  which uses the arc closure algorithm  for the analysis of knotted structures in  linear chains\cite{tubiana2011multiscale,tubiana2011probing}.
Importantly, the average fragment $l_h$ that is part of a knot/knots exhibits a non-monotonic behavior with  frequency  for various pore radii, as depicted in Fig.~\ref{fig:knots}-a. The $l_h$ shows the  extent of knotted length feasible within a polymer under field. In the equilibrium or at higher frequency limit (like for $R_p=20$, $\nu \ge  5\times 10^{-3}$)  we found $l_h=0$, which indicates that the structure is devoid of any such knots. While for the intermediate frequencies, the value of $l_h$ sharply grows even reaching beyond $0.5$, suggesting an enhanced knotting favourability, where a large portion of the chain is topologically  entangled.  
 The attainability of the chain self-entanglement to such a wide extent is ensued from the oscillatory nature of the inducing field  leading to  crumbled structures with low structural expansion.
Further, as we go to the DC limit ($\nu<<10^{-4}$, $l_h$ is shown with arrows), the fraction constituting the knot falls off dramatically, but exhibits a non-zero $l_h$ (see Fig.~\ref{fig:knots}-a).  In DC, the formation of a well taut and stretched tadpole constitutes a tightly knotted head (see Fig.~\ref{fig:knots}-b-ii), resulting in  smaller values of $l_h$, contrary to the loosely formed knotted fragments observed  at intermediate frequencies in AC (see Fig.~\ref{fig:knots}-b-i). This can be understood by considering a piece of rope with a simple knot tied on it, such that if we pull the chain from its terminus, the knotted part will become tight and localized.  The top row (i) and (ii)  of Fig.~\ref{fig:knots}-b,  shows the native chain structures formed under AC field (knotted portion in orange), while the bottom row (iii) and (iv) shows the corresponding tight and localized knot formed after stretching the field favoured native structures at its extremities. 
\begin{figure}	
\includegraphics[width=\linewidth]{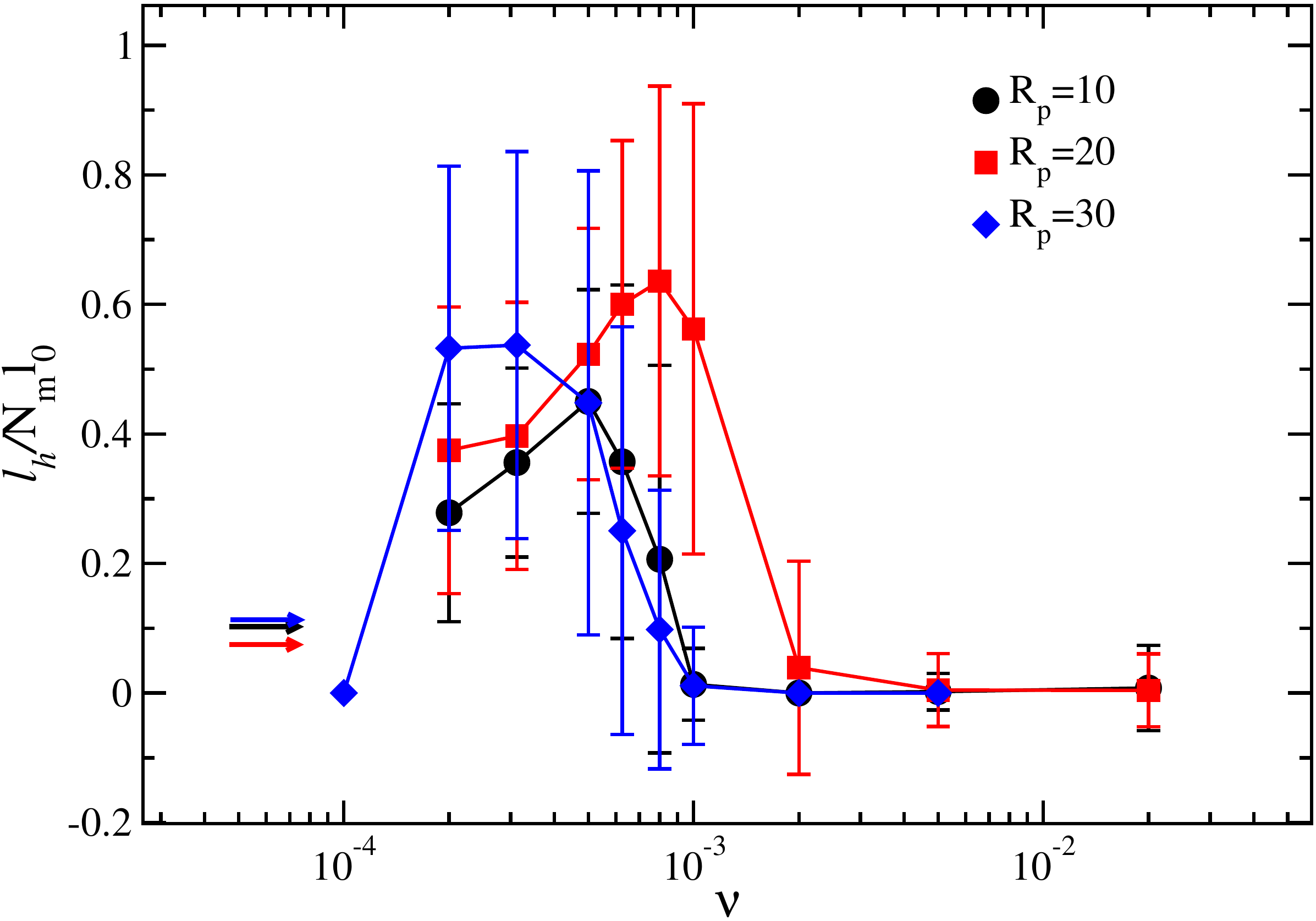}\\
\includegraphics[width=\linewidth]{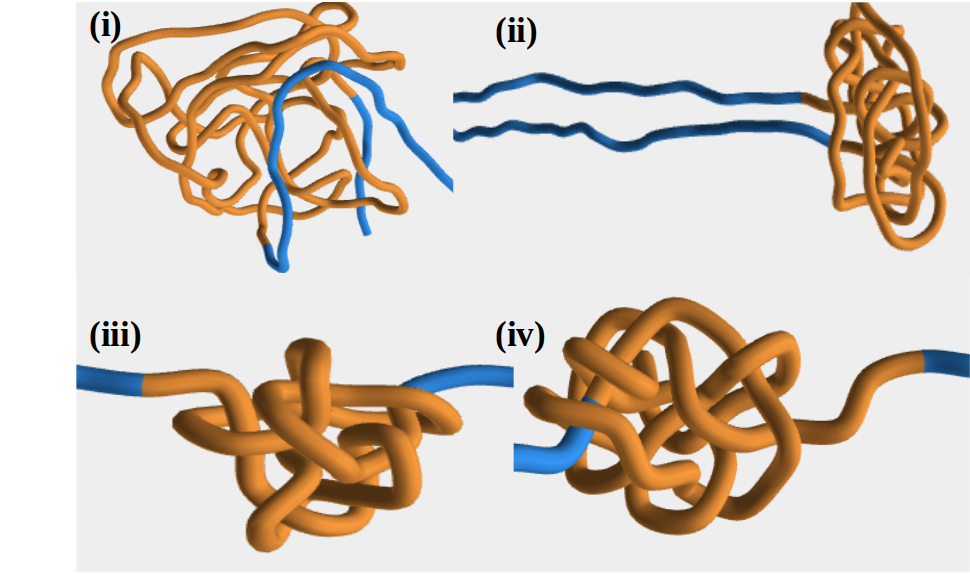}\\
   \caption{The fraction of chain length $l_h$ that constitutes a knot as a function $\nu$ at field strength $G=2.0$, for different pore radii obtained using \textit{Kymoknot} software\cite{tubiana2018kymoknot}.  The error bar depicts the standard deviation at each frequencies.b) Top row depicts few snapshots of the chain conformations obtained under field, with the knotted portion highlighted and, bottom row corresponds to the tight and localized knots obtained after stretching  respective conformations at its ends.   }
    \label{fig:knots}
\end{figure}



{\cblue 
}


 \section{ Discussion } 
 We have presented the phenomenon of chain collapse to self entangled structures orchestrated by direct and oscillatory field using a generic polymer model. Under any direct force, the generation of recirculating hydrodynamic flow fields paves way to tadpoles structures, but the  imposition of spatial constriction profoundly influences the intra-chain intertwining, leading to field compressed  structures even in longer chains.  For larger bending rigidity, this collapse  exhibits a remarkable non-monotonic response, where  the compression is dictated by the competition between the bending cost and stretching force.
 Apart from the confinement and chain rigidity, nature of the jostling force also brings fascinatingly complex dynamics into picture\cite{belmonte2001dynamic,soh2019self,ben2001knots} . Following this, we elucidate that a semi-flexible chain under an oscillating field exhibits  remarkable compression in certain frequency window. The AC field orchestrated compression is attainable across a wide range of confinements (including bulk), bending rigidities and field strengths, where chain simply stretches  under DC. The field switching under AC captures the initial recirculating flow induced compaction, forbidding proper tail extension, such that the chain essentially gets arrested in a collapsed state with fluctuating ends. 
 
 Within the cellular structure, self organisation and  dynamical instabilities causes  mechanical oscillations\cite{kruse2005oscillations}. These periodic jostling forces  might profoundly influence the conformations of the biopolymer, mediated via the cellular fluid. Further, dielectrophoresis (DEP) experiments\cite{viefhues2017dna,asbury2002trapping,Zhou_PRL_2011} generally involves high fields of alternating nature, where this oscillatory field is leveraged  for more controlled manipulation and efficient separation schemes for DNA and other polymers. 

}


\text{\it Acknowledgments: }
This work received financial support from the DST SERB Grant No. CRG/2020/000661. High-performance computing facility provided  at IISER Bhopal and Paramshivay NSM facility at IIT-BHU are also acknowledged.   

\end{document}